\documentclass[aps,pre,twocolumn,showpacs,color,longbibliography] {revtex4-1}
\usepackage{graphicx} 
\usepackage{color}
\usepackage{mwe}
\usepackage{amssymb,amsfonts,amsmath}
\usepackage[normalem]{ulem} 
\definecolor{g-blue}{rgb}{0.83,0.95,1}
\definecolor{g-yellow}{rgb}{1,1,0.7}
\definecolor{g-green}{rgb}{0.9,1,0.9}
\definecolor{green}{rgb}{0,0.6,0}
\definecolor{cyan}{rgb}{0,0.7,0.7} 
\definecolor{grey}{rgb}{0.4 ,0.4 ,0.4 }
\definecolor{brown}{rgb}{0.6 ,0  ,0.8 }

\def\g-blue#1{\textcolor{g-blue}{#1}}

\def\red#1{\textcolor{red}{#1}}

\begin{document}

\title{Anomalous correlators in nonlinear dispersive wave systems}

\author{ Joseph Zaleski$\ ^1$, Miguel  Onorato$\ ^2$ and Yuri V Lvov
  $\ ^1$\\ 
  $^1$ {Department of Mathematical Sciences, Rensselaer
    Polytechnic Institute, Troy, New York 12180, USA},\\
$\ ^2${Dip. di Fisica, Universit\`{a} di Torino
    and INFN, Sezione di Torino, Via P. Giuria, 1 - Torino, 10125,
    Italy},
}

\begin{abstract}
  We show that Hamiltonian nonlinear dispersive wave systems with
  cubic nonlinearity and random initial data develop, during their
  evolution, anomalous correlators. These are responsible for the
  appearance of ``ghost'' excitations, i.e. those characterized by
  negative frequencies, in addition to the positive ones predicted by
  the linear dispersion relation. We use generalization of the Wick's
  decomposition and the wave turbulence theory to explain
  theoretically the existence of anomalous correlators.  We test our
  theory on the celebrated $\beta$-Fermi-Pasta-Ulam-Tsingou chain and
  show that numerically measured values of the anomalous correlators
  agree, in the weakly nonlinear regime, with our analytical
  predictions.  We also predict that similar phenomena will occur in
  other nonlinear systems dominated by nonlinear interactions,
  including surface gravity waves. Our results pave the road to study
  phase correlations in the Fourier space for weakly nonlinear dispersive
  wave systems.
\end{abstract}

\maketitle 

\section{Introduction}
Wave Turbulence theory has led to successful predictions on the wave
spectrum in many fields of physics
\cite{falkovich1992kolmogorov,nazarenko2011wave}.  In this framework
the system is represented as a superposition of a large number of
weakly interacting waves with the complex normal variables
$a_k=a(k,t)$. In its essence, the classical Wave Turbulence theory is
a perturbation expansion in the amplitude $a_k$ of the nonlinearity,
yielding, at the leading order, to a system of quasi-linear waves
whose amplitudes are slowly modulated by resonant nonlinear
interactions~\cite{falkovich1992kolmogorov,nazarenko2011wave,Newell,N1,Ben,K}.
This modulation leads to a redistribution of the spectral energy
density among length-scales, and is described by a wave kinetic
equation.  One way to derive the wave kinetic equation is to use the
random phase and amplitude approach developed in
\cite{nazarenko2011wave,X1,X2}.  The initial state of the system can
be always prepared so that the assumption of random phases and
amplitudes is true.  Whether the phases remain random in the evolution
of the system has been an issue of intense discussions. In Wave
Turbulence theory, the standard object to look at is the second-order
correlator, $\langle a_k(t)a^*_l(t)\rangle$, where
$\langle\dots\rangle$ is an average over an ensemble of initial
conditions with different random phases and amplitudes.  As will be
clear later on, under the homogeneity assumption, the second order
correlator is related to the wave action spectral density function,
i.e. the {\it wave spectrum}, $n_k=n(k,t)$.  However, one should note
that the complex normal variable, as defined in the the Wave
Turbulence theory, is a complex function also in physical
space. Therefore, the second-order statistics are not fully determined
by the above correlator.  The so called ``anomalous correlator'',
$\langle a_k(t)a_l(t)\rangle$, see \cite{l2012wave,zakharov1975spin},
needs also to be computed.  Under the hypothesis of homogeneity, will
be related to the {\it anomalous spectrum}, $m_k=m(k,t)$, to be
defined in the next Section. Indeed, if phases are totally random,
this quantity would be zero. We show that, in the nonlinear evolution
of the system, this is not the case.  Far from it, this quantity is
strongly nonzero and, in the limit of weak nonlinearity, we predict
analytically and verify numerically its value.

{ Our ideas are based on the extension of the Wave
  Turbulence Theory to include these anomalous correlators.  Notably,
  conventional Wave Turbulence Theory has been successful in the
  understanding of the spectral energy transfer in complex wave
  systems such as the ocean \cite{janssenbook04}, optics
  \cite{Picozzi2014} and Bose-Einstein condensates \cite{Proment2012},
  one dimensional chains \cite{PNAS}, and magnets \cite{LVOVBOOK}.
  Analogously, anomalous correlators first appeared in the well known
  Bardeen, Cooper, and Schriffer (BCS) theory of superconductivity
  \cite{BCSTheory}.  Subsequently, anomalous correlators have been
  studied in S--theory \cite{zakharov1975spin,l2012wave}.

Recently, anomalous correlations were shown to play an important role
in explaining numerical observations of nondecaying oscillations
around a steady state in a turbulence--condensate system modeled by
the Nonlinear Schr\"odinger equation
\cite{NLSPhaseCorr,NLSsim1,NLSsim2}. Such oscillations, corresponding
to a fraction of the wave action being periodically converted from the
condensate to the turbulent part of the spectrum, were shown to be
directly due to phase coherence \cite{NLSPhaseCorr}.  In
\cite{guasoni2017incoherent} a system of Coupled Nonlinear
Schr\"odinger equations has been considered and specific attention was
focussed on the phenomena of recurrence of incoherent waves observed
in the early stages of the dynamics. The authors derived a variant of
the kinetic equation which includes anomalous correlators; the
peculiarity of such an equation is that it is capable of describing
properly the recurrence phenomena observed in the simulations.  }

One of the main tools used to derive the theory is the Wick's
contraction rule that allows one to split higher order correlators as
a sum of products of second order correlators, plus cumulants.  To
explain analytically the existence of the anomalous correlators, it is
necessary to use the more general form of the Wick's decomposition,
namely the form that allows anomalous correlators. We then demonstrate
that the anomalous correlators are responsible for creating the
``ghost waves'', i.e. the waves with the frequency equal to the
negative of the frequency predicted by the linear dispersion
relationship.  These ideas are tested on a simple, but non trivial,
system, i.e. the $\beta$-Fermi-Pasta-Ulam-Tsingou (FPUT) chain.  The
chain model was introduced in the fifties to study the thermal
equipartition in crystals \cite{fermi1955studies}; it consists of $N$
identical masses, each one connected by a nonlinear
{{string}}; the elastic force can be expressed as a power
series in the { displacement from equilibrium.} Fermi,
Pasta, Ulam and Tsingou integrated numerically the equations of motion
and conjectured that, after many iterations, the system would exhibit
a thermalization, i.e. a state in which the influence of the initial
modes disappears and the system becomes random, with all modes excited
equally (equipartition of energy) on average. Successful predictions
on the time scale of equipartition have been recently obtained in
\cite{PNAS, pistone2018thermalization,lvov2018double} using the Wave
Turbulence approach. In this paper we perform extensive numerical
simulations with initial random data and look all at the possible
excitations, once a thermalized state has been reached.  This is all
done by analyzing the spatial -temporal $(k-\Omega)$ spectrum,
i.e. the square of the space-time Fourier transform of the wave
amplitudes.  Analyses of the effective dispersion relation in the
nonlinear system is a well known and widely used theoretical and
numerical tool, see for example \cite{kovacicPNAS}.

We give numerical evidence that in addition to the ``normal'' waves
with frequency $\omega$ predicted by the linear dispersion relation
for wave number $k$, there are the ``ghost'' excitations with the
negative frequencies. Our theoretical analysis reveals that the origin
of those ``ghost'' excitations resides on the nonzero values of the
second-order anomalous correlator.

\section{The Model}
The theory that we develop hereafter applies to any system with cubic
nonlinearity. Examples of such systems among others, include deep
water surface gravity waves ~\cite{zakharov68}, Nonlinear Klein Gordon
\cite{pistone2018thermalization}, $\beta$-Fermi-Pasta-Ulam-Tsingou
chain. In normal variables $a_k$ the Hamiltonian of these systems
assumes the canonical form:
\begin{equation}
\begin{split}
&H=\sum_k{\omega_k |a_k|^2}+\sum_{k_1,k_2,k_3,k_4}
\big[
(T^{(1)}_{1234} a_1^*a_2a_3a_4+c.c)\delta_{1}^{234} \\
&+\frac{1}{2}T^{(2)}_{1234} a_1^*a_2^*a_3a_4\delta_{12}^{34}
+ \frac{1}{4}T^{(4)}_{1234} (a_1^*a_2^*a_3^*a_4^*+c.c)\delta_{1234}\big],
\end{split}
\end{equation}
where $\omega_k=\omega(k)$ are the positive frequencies associated to
the wave numbers via the dispersion relation, $T^{(i)}_{1234}$ are
coefficients that depend on the problem considered and satisfy
specific symmetries for the system to be Hamiltonian, $c.c.$ implies
complex conjugation, $a_j=a(k_j,t)$ are the complex normal variables,
$\delta_{ij}^{lm}=\delta(k_i+k_j-k_l-k_m)$ is the Kronecker Delta.  We
assume that the only resonant interactions possible are the ones for
which the following two relations are satisfied for a set of wave
numbers
\begin{equation}
k_1+k_2=k_3+k_4,\;\;\; \omega(k_1)+\omega(k_2)=\omega(k_3)+\omega(k_4).
\end{equation}

With the objective of presenting some comparison with numerical
simulations, out of many physical systems described by the above
Hamiltonian, we select a simple one dimensional system, the
$\beta$-Fermi-Pasta-Ulam-Tsingou chain. {Modeling a
  vibrating string,} this problem {consists of a system of
  $N$ identical particles connected locally to each other by a
  nonlinear oscillator}. In the physical space the displacements with
respect to the equilibrium position $q_j(t)$ and their momenta
$p_j(t)$, the Hamiltonian takes the following form:
\begin{equation}
H=H_2+H_4
\end{equation}
with 
\begin{equation}
\begin{split}
&H_2=\sum\limits_{j=1}^N\left(\frac{1}{2 }p_j^2+\frac{1}{2}(q_j-q_{j+1})^2\right),
\\
&H_4=\frac{\beta}{4}\sum\limits_{j=1}^N(q_j-q_{j+1})^4.
\end{split}
\label{H_FPU}
\end{equation}
$\beta$ is the nonlinear spring coefficient (without loss of
generality, we have set the masses and the linear spring constant
equal to 1).  The Newton's law in physical space is given by:
\begin{equation}
 \ddot{q}_j=(q_{j+1}+q_{j-1}-2q_j)+
\beta\big[(q_{j+1}-q_j)^3-(q_j-q_{j-1})^3\big].
\label{eq:eqmotion}
\end{equation}

We assume periodic boundary condition; our approach is developed in
Fourier space and the following definitions of the direct and inverse
Discrete Fourier Transforms are adopted:
\begin{equation}
Q_k=\frac{1}{N}\sum_{j=0}^{N-1} q_j e^{-i 2\pi k
	j/N},\;q_j=\sum_{k=-N/2+1}^{N/2} Q_k e^{ i 2\pi j k/N}, \label{DFT}
\end{equation}
where $k$ are discrete wave numbers and $Q_k$ are the Fourier
amplitudes. The displacement  $q_j$ and
 momentum $p_j$ of the $j$ particle are linked by canonically
 conjugated Hamilton equations
\begin{equation}
\dot p_j = -\frac{\partial H}{\partial q_j}, \dot q_j
    =\frac{\partial H}{\partial p_j}.\nonumber
\end{equation}
    
We then perform the Fourier transformation to Fourier images of
position and momenta, and then additional canonical transformation to
complex amplitude $a_k$ given by
\begin{equation}
a_k=\frac{1}{\sqrt{2 \omega_k}}(\omega_k Q_k+i P_k),\label{NormalMode}
\end{equation}
%
where $\omega_k=2|\sin(\pi k/{N})| > 0$ and $Q_k$ and $P_k$ are the
Fourier amplitudes of $q_j$ and $p_j$, respectively. In terms of $a_k$ the equation of motion 
reads, see \cite{bustamante2019exact}:
\begin{equation}
\begin{split}
&i\frac{d a_1}{d t}
=\omega_{k_1} a_{1}
+ \sum_{k_2,k_3,k_4} \big(
T^{(1)}_{1234} a_2a_3a_4\delta_{1}^{234} 
+T^{(2)}_{1234} a_2^*a_3a_4\delta_{12}^{34}+\\
&+ T^{(3)}_{1234}a_2^*a_3^*a_4\delta_{123}^{4}
+ T^{(4)}_{1234} a_2^*a_3^*a_4^*\delta_{1234}\big),
 \label{eq:fourwaveint}
\end{split}
\end{equation}
where all wave numbers $k_2, k_3$ and $k_4$ are summed from $0$ to
$N-1$ and $\delta_{ab..}^{cd..}=\delta(k_a+k_b+...-k_c-k_d-...)$ is
the generalized Kronecker Delta that accounts for a periodic Fourier
space, i.e. its value is one when the argument is equal to 0
(\text{mod} $N$). The matrix elements $T^{(1)}_{1234}$,
$T^{(2)}_{1234}$, $ T^{(3)}_{1234}$, prescribe the strength of
interactions of wave numbers $k_1$, $k_2$, $k_3$ and $k_4$. Their
values \eqref{matrixelements} are given in Appendix A.
\subsection{The $(k-\Omega)$ spectrum}
The main statistical object discussed in this paper is the wave
number-frequency $(k-\Omega)$ spectrum.  Starting from the complex
amplitude $a(k,t)$ we take the Fourier transform in time so that we
get $a(k,\Omega)$; Under the hypothesis of homogeneous and stationary
conditions, the second-order $(k-\Omega)$ correlator takes the
following form
 \begin{equation}
 \langle a(k_i,\Omega_p) a(k_j,\Omega_q)^*\rangle=N(k_i,\Omega_p)\delta(k_i-k_j)\delta(\Omega_p-\Omega_q),
 \label{eq:kospectrum}
 \end{equation}
 where $ \langle ....\rangle$ implies averages over initial conditions
 with different random phases. 
$N(k,\Omega)$ is the $(k-\Omega)$ spectrum 
 defined as follows:
 \begin{equation}
N^{(a)}(k,\Omega)=\frac{1}{2\pi}\frac{1}{N}\int_{-\infty}^{+\infty}\sum_{l=1}^{N} 
R(l, \tau)\mathrm{e}^{-i 2 \pi k l /N} \mathrm{e}^{-i\Omega \tau}d\tau,
 \end{equation}
 with $R(l, \tau)=\langle a_j(t)^* a_{j+l}(t+\tau)\rangle$ is the
 space-time autocorrelation function.\\ {\it The linear $(k-\Omega)$
   spectrum - } Before diving into the nonlinear dynamics, we discuss
 the predictions in the linear regime. Therefore, we start by
 neglecting the nonlinearity in equation (\ref{eq:fourwaveint}) and
 find the solution in the form
 \begin{equation}
 \begin{split}
 a_{k}(t)=a_{k}(t_0) e^{-i \omega_{k} t}.
 \end{split}
 \end{equation}
 where $t_0$ is a time at which the solution is known or an initial condition.
 We then take Fourier transform in time
 \begin{equation}
 \begin{split}
 a(k,\Omega)=a(k,t_0) \delta(\Omega-\omega_k)
 \end{split}
 \end{equation}
After multiplication by its complex conjugate and taking averages over
different realizations with the same statistics, we get:
 \begin{equation}
\label{eq:komega}
N^{(a)}(k,\Omega)=n^{(a)}(k,t_0)\delta(\Omega-\omega_k),
\end{equation}
 where $n^{(a)}(k,t_0)$ is the standard wave spectrum at time $t_0$
 related to the second-order correlator as
 \begin{equation}
 \langle a(k_i,t_0) a(k_i,t_0)^*\rangle=n^{(a)}(k_i,t_0)\delta(k_i-k_j).
   \end{equation}
   and defined via the autocorrelation function as 
   \begin{equation}
  n^{(a)}(k_i,t_0)= \frac{1}{N}\sum_l \langle a_j(t_0) a_{j+l}(t_0)^*\rangle e^{-i 2\pi k l/N}.
\end{equation}
 In the linear regime $n^{(a)}(k_i,t_0)$ does not evolve in time. 
 
 Equation (\ref{eq:komega})
   implies that in the linear case the $(k-\Omega)$ spectrum is
   different from zero only for those values of $\Omega$ and $k$ for
   which the dispersion relation is satisfied.  Note that in this
   formulation $\omega_k$ is defined as a positive quantity;
   therefore, only the positive branch of the dispersion relation
   curve appears in the linear regime.

\subsection{Numerical results for the $(k-\Omega)$ spectrum}
\begin{figure*}
\centering
\includegraphics[width=0.9\linewidth]{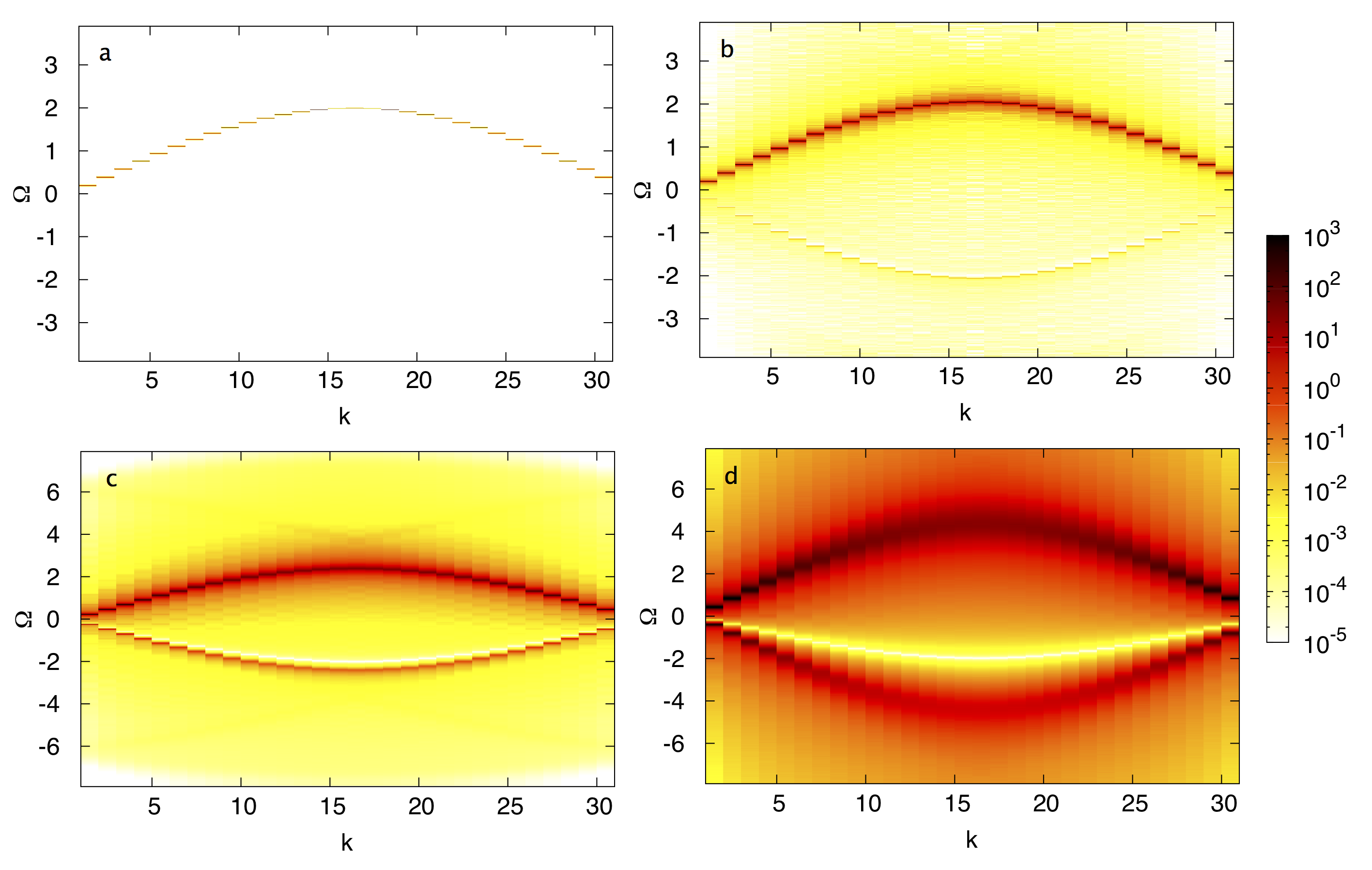}\hfil
 \caption{$(k-\Omega)$ spectrum, $N^{(a)}(k,\Omega)$, for different values of $\epsilon$: 
 $(a)$ $\epsilon=0$, 
$(b)$ $\epsilon=0.0089$
$(c)$ $\epsilon=0.089$
   $(d)$ $\epsilon=1.12$.
In the linear case, $(a)$, the $N^{(a)}(k,\Omega)$ is different from 0
only when the frequency $\Omega$ matches the linear dispersion
relation. As the nonlinearity is increased, $(b-d)$, a frequency
shift, a broadening of the frequencies and a lower branch less intense
than the upper one are visible. Waves with negative frequencies are
named ``ghost'' excitations.}
\label{komegaplots}
\end{figure*}

We now test the predictions from equation (\ref{eq:komega}) both in
the linear regime and observe what happens to it in the nonlinear
regime.  We perform numerical simulations of the equations
(\ref{eq:eqmotion}) using a symplectic algorithm, see
\cite{yoshida1990construction}. We use 32 particles in the
simulations; such choice is completely uninfluential for the results
presented below.  In the linear regime, we just prescribe a
thermalized spectrum with some initial random phases of the wave
amplitudes $a_k$ and evolve the system in time up to a desired final
time; a Fourier Transform in time is then taken to build the
$(k-\Omega)$ spectrum. In the nonlinear regime we perform long
simulations up to a thermalized spectrum. For a given nonlinearity,
1000 realizations characterized by different random phases are made
and ensemble averages are considered to compute the $(k-\Omega)$
spectrum. All simulations have the same initial linear energy and,
from an operative point of view, the only difference between them is
the value of $\beta$.  To characterize the strength of the
nonlinearity, we use the following ratio between nonlinear and linear
Hamiltonians at the beginning of each simulation:
\begin{equation}
\epsilon=\frac{H_4}{H_2}\propto \beta
\end{equation}
Results are shown in Figure \ref{komegaplots}, where,  for
different values of the nonlinear parameter $\epsilon$, the spectrum
$N^{(a)}(k,\Omega)$ is plotted using a colored logarithmic scale.  We
first focus our attention on the linear regime, $\epsilon=0$: results
are shown in Figure \ref{komegaplots}(a).
\begin{figure}
\centering
\includegraphics[width=0.8\linewidth]{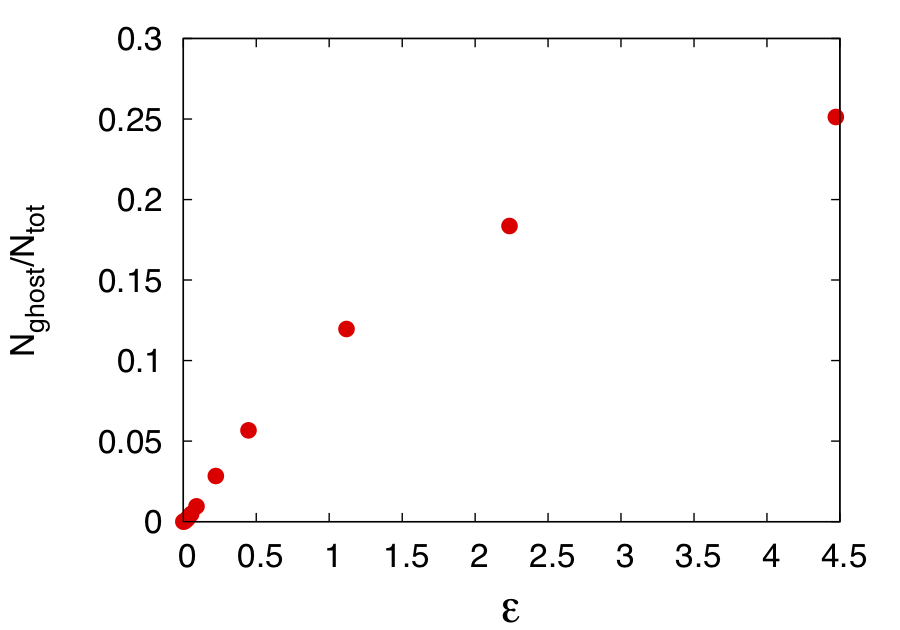}
 \caption{Ratio between the number of ``ghost'' excitations, $\text{N}_{\text{ghost}}$, over the total number of waves, $\text{N}_{\text{tot}}$,as a function of the nonlinearity.}
 \label{ratio}
\end{figure}
As well predicted by the theory, the plot shows 
dots in the positive frequency plane, where the frequencies $\Omega$
and wave numbers $k$ satisfy the linear dispersion curve
$\omega_k$.
Increasing the nonlinearity, Figures \ref{komegaplots} -(b,c,d),
two well known effects appears: the first one is a shift of the
frequencies, due to nonlinearity (this is more evident in Figures
\ref{komegaplots} -(c,d) where the frequency scale in the vertical
axes has been changed). The second one is the broadening of the
frequencies; this is related to the fact that the amplitude for each
wave number is not constant in time; therefore, the
amplitude-dependent frequencies are not constant in time and they
oscillate around a mean value with some fluctuations.  Those results
are well understood, at least in the weakly nonlinear regime, and can
be predicted using Wave Turbulence tools, see
\cite{nazarenko2011wave,lvov2018double}. Besides these two effects,
starting from Figure \ref{komegaplots}-(b), the presence of a
lower branch, whose intensity is much less than the upper one, starts
to be visible. The lower curve becomes more important and, when the
nonlinearity is of order one, is of the same order of magnitude of the
upper one.
The total number of waves in the simulation, $\text{N}_{\text{tot}}$,  is given by the integral over $\Omega$ and the sum over all $k$ of the function $N^{(a)}(k,\Omega)$. 
In the weakly nonlinear regime, $\text{N}_{\text{tot}}$ is an
adiabatic invariant of the equation of motion (\ref{eq:eqmotion}); the
plot highlights the existence of waves with negative frequencies,
which will be named ``ghost'' excitations. One of the scopes of the
present paper is the understanding of the origin of such waves.
Before entering into the discussion, we show in Figure \ref{ratio}
the ratio of ``ghost'' excitations, $\text{N}_{\text{ghost}}$,
i.e. $N^{(a)}(k,\Omega)$ integrated over negative frequencies and
summed over all wave numbers, divided by the total number of waves,
$\text{N}_{\text{tot}}$. As can be seen from the plot, there is a
monotonic growth of the ``ghost'' waves that, for very large
nonlinearity, can reach values up to 25$\%$ of the total number.
\section{Anomalous correlators} 
To explain the presence of ``ghost'' excitations, we introduce the so called second-order anomalous correlator
\cite{l2012wave,zakharov1975spin, guasoni2017incoherent}:
\begin{equation}
\label{eq:anomal1}
\langle a_{k}(t)a_{j}(t)\rangle= m_{k}(t) \delta(k+j),
\end{equation}
with the {\it anomalous spectrum} defined as
\begin{equation}
\label{eq:anomal2}
m_k^{(a)}(t)=\frac{1}{N}\sum_l \langle a_j a_{j+l}\rangle e^{-i 2\pi k l/N}.
\end{equation}

Similarly, we also introduce the second-order $(k-\Omega)$ anomalous correlator:
\begin{equation}
\label{eq:anomal3}
\langle   a(k_i,\Omega_l)  a(k_j,\Omega_m)\rangle=
M^{(a)}(k_i,\Omega_l)\delta(k_i+k_j)\delta(\Omega_l+\Omega_m),
\end{equation}
where
 \begin{equation}
M^{(a)}(k,\Omega)=\frac{1}{2\pi}\frac{1}{N }\int_{-\infty}^{+\infty}\sum_{l=1}^{N} 
S(l, \tau)\mathrm{e}^{-i 2 \pi kl/N} \mathrm{e}^{-i\Omega \tau}d\tau
 \end{equation}
and $S(l, \tau)=\langle a_j(t) a_{j+l}(t+\tau)\rangle$. The presence  in equations
  (\ref{eq:anomal1}) and (\ref{eq:anomal3}) of the Kronecker
$\delta$ over wave numbers and the Dirac $\delta$ over frequency, are
related to the hypothesis of  statistical homogeneity and stationarity, respectively. 
Note that  $M^{(a)}(k,\Omega)$
is not the Fourier transform in time of $m_k^{(a)}(t)$ and in general 
both can be complex functions. 
\begin{figure*}
\centering
\includegraphics[width=0.9\linewidth]{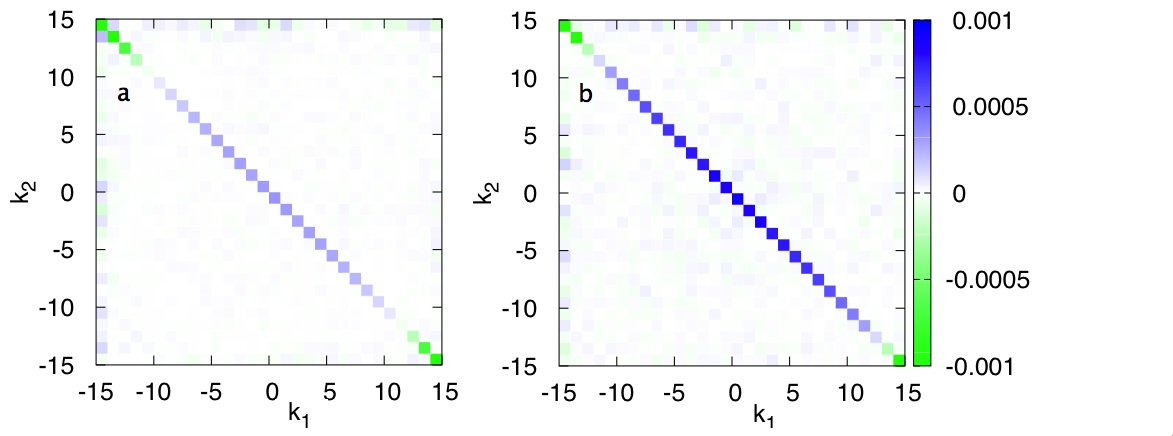}
 \caption{The real part of the second-order anomalous  correlator, 
  ${\text Re}[\langle a_{k_1} a_{k_2}\rangle]$, for 
 (a) $\epsilon=0.0089$,
 (b) $\epsilon=1.12$. A diagonal contribution corresponding to 
 $k_2=-k_1$ is evident in both figures. As the nonlinearity is increased, 
 the contribution becomes larger.}
 \label{figm}
\end{figure*}
To verify numerically that the anomalous correlator is indeed nonzero,
we measure numerically the real part of the second-order correlator
$\langle a_{k_i}(t)a_{k_j}(t)\rangle$ as a function of $k_1$ and
$k_2$. Results are plotted in Figure \ref{figm} where we show the results of
two numerical simulations characterized by two different values of the
nonlinear parameter, (a) $\epsilon=0.0089$ and (b) $\epsilon=1.12$.
In both cases, a diagonal contribution is visible, pointing out the
existence of anomalous correlators in the $\beta$-FPUT model.  

{\it
  Generalization of the Wick's decomposition - } Using \eqref{eq:anomal1}, it is
    straightforward to extend the Wick's decomposition by taking
    into account the anomalous correlators, as done in
    \cite{LVOVBOOK}:
\begin{equation}
\begin{split}
&\langle a_k^* a^*_l a_p a_n \rangle
= n_k n_l(\delta^k_p\delta ^l_n + \delta^k_n\delta ^l_p) +
m^*_k m_p\delta_{kl}  \delta _{pn}, \\
&\langle   a_k^* a_l a_p a_n\rangle =
n_k m_p(\delta_k^l \delta _{pn}+ \delta_k^n \delta_{lp})+ n_k \delta^p_k m_l\delta_{nl},\\
&\langle   a_k a_l a_p a_n\rangle = m_k m_l (\delta_{kp}\delta_{ln}+\delta_{kl}\delta_{pn}+\delta_{kl}\delta_{pn}).
\label{eq:wick}
\end{split}
\end{equation}
The above relations will be fundamental for making a natural closure
of the moments when calculating analytically the $(k-\Omega)$
spectrum.

\begin{figure*}
\centering
\includegraphics[width=0.9\linewidth]{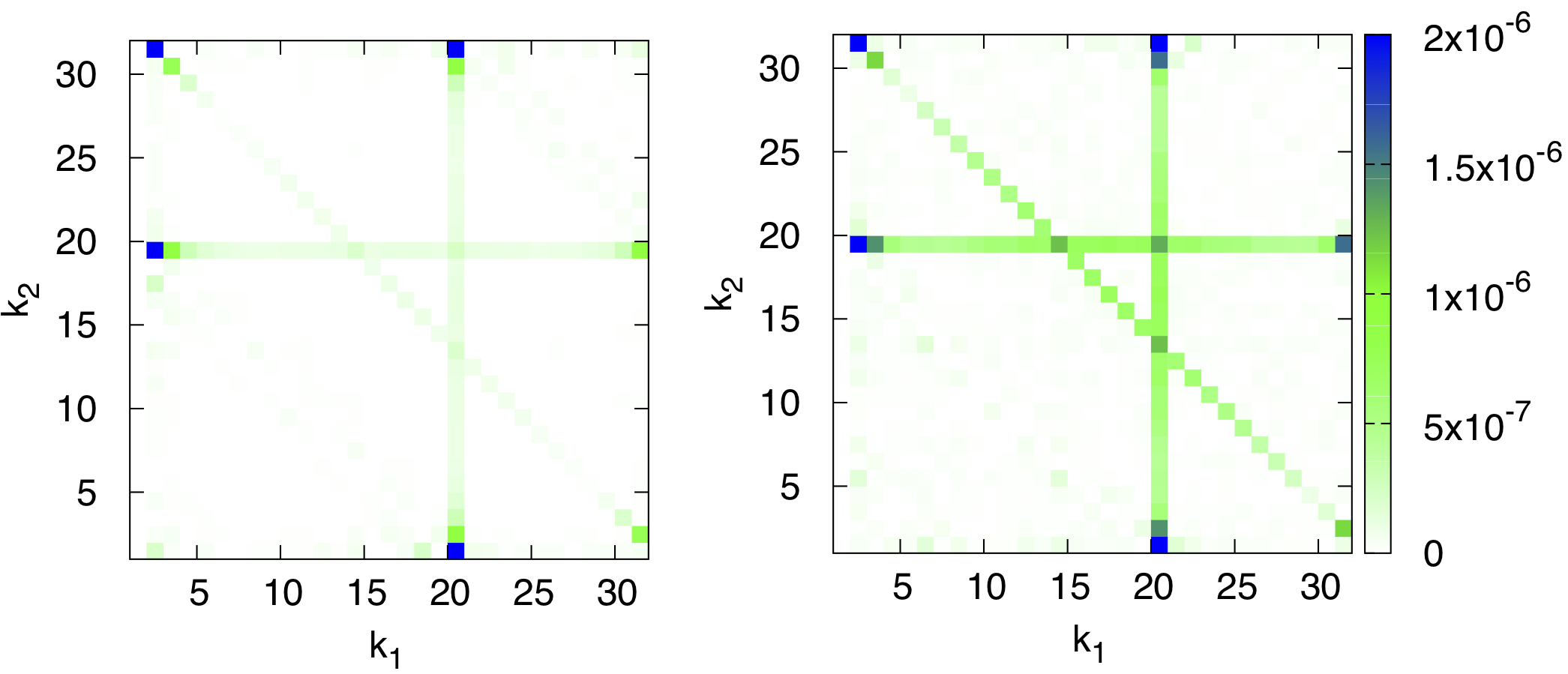}\hfil
\caption{
  Fourth-order correlator  
 $|{\text Re}[\langle a_{k_1} a_{k_2}a_{k_3}^*a_{k_1+k_2-k_3}^*\rangle]|$ with $k_3$=20.
 (a) $\epsilon=0.0089$,
 (b) $\epsilon=1.12$. Different horizontal, vertical and diagonal  lines are visible. 
 Horizontal and vertical lines corresponds to trivial resonances: 
$ k_1=k_3$, vertical line; $ k_2=k_3$, horizontal line; 
$ k_2=-k_1+N$, diagonal line. The latter line corresponds to the presence of 
an anomalous second-order correlator. The intensity of the lines is larger for 
larger nonlinearity.
 }
 \label{fig:corr4}
\end{figure*}

In Figure \ref{fig:corr4}, we find further evidence justifying this decomposition by plotting the real part of the fourth-order
correlator $\langle a_{k_1} a_{k_2}a_{k_3}^*a_{k_1+k_2-k_3}^*\rangle$
with $k_3=20$, computed from numerical simulations for (a)
$\epsilon=0.0089$ and (b) $\epsilon=1.12$. The diagonal lines in both
figures, highlighting the contribution from the second-order anomalous
correlator, are noticeable. The vertical and horizontal lines
correspond to the trivial resonances in which two wave numbers are
equal (\text{mod} $N$).

\subsection{Theoretical prediction for the anomalous correlator in the weakly nonlinear regime}
A key step for the development of a theory for the anomalous
correlator is the change of variable (near identity transformation)
which allows one to remove {\it bound} modes, i.e. those modes that
are phase locked to {\it free} modes and do not obey the linear
dispersion relation. The procedure is well known in Hamiltonian
mechanics and well documented for example in
\cite{falkovich1992kolmogorov}. We accomplish this via the following
canonical    transformation  from variable $a_k(t)$ to
    $b_k(t)$
\begin{equation}
\begin{split}
&a_1=b_1+\sum\limits_{k_2,k_3,k_4}\big[B^{(1)}_{1234}b_2b_3b_4 \delta^{234}_{1} +
B^{(3)}_{1234}  b_2^*b^*_3b_4\delta^4_{123}+ \\
&+B^{(4)}_{1234}b^*_2b^*_3b^*_4\delta_{1234}\big],
\label{CanonicalTrans}
\end{split}
\end{equation}
with the coefficients $B^{(i)}_{1234}$ selected in such a way to remove
non resonant terms in the original Hamiltonian \cite{krasitskii1994reduced}.
{\red Their values are given in Appendix A.}

The transformation is asymptotic in the sense that the small amplitude
approximation is made and the terms in the sums on the right hand
side are much smaller than the leading order term $b_1$. The evolution
equation for variable $b_k(t)$ contains resonant interactions and take
the following standard form:
\begin{equation}
\begin{split}
i\frac{d b_1}{\partial t}=\omega_1 b_1 + \sum_{k_ 2, k_3,k_4}  T^{(2)}_{1234}
b_2^*b_3b_4\delta_{12}^{34} +{\text{h.o.t.}}
\nonumber
\label{eq:zakh}
\end{split}
\end{equation}
where higher order terms arising from the transformation have been
neglected.  

Using the transformation (\ref{CanonicalTrans})
and the generalized Wick's decomposition (\ref{eq:wick}), we can
now build the time averaged anomalous spectrum (for details, see Appendix B):
\begin{equation}
\begin{split}
 & \langle m_k^{(a)}(t)\rangle_t= 2  \left(n_k^{(a)}+n_{-k}^{(a)}\right)\sum_{j} B^{(3)}_{k,-k,j,j}n_{j}^{(a)},
\end{split}
\label{mkEq}
\end{equation}
where $\langle ... \rangle_t$ implies averaging over time. For the $\beta$--FPUT system in thermal equilibrium, where $n^{(a)}_k =T/\omega_k$ with $T$ constant, \eqref{mkEq} reduces to
\begin{equation}
    \omega_k|\langle m_k^{(a)}(t)\rangle_t|=\frac{3NT^2\beta}{2}.
    \label{mkThe}
\end{equation}

\begin{figure} [h] 
\label{fig:T}
\centering
\includegraphics[width=1.\linewidth]{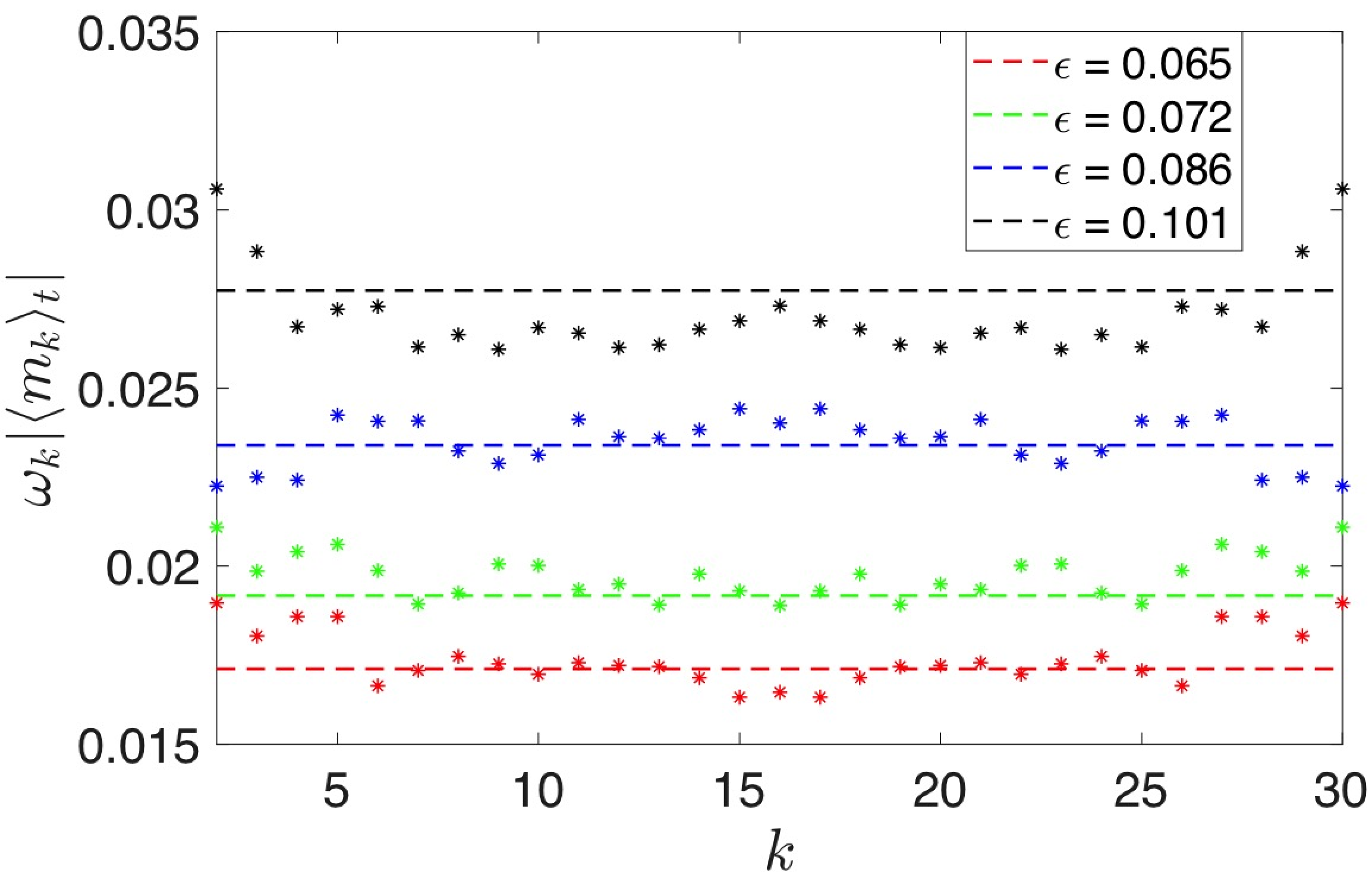}
\caption{
Comparison of anomalous spectrum 
 $\omega_k |\langle m^{(a)}_k(t)\rangle_t|$ as observed in numerical simulations (dots) with theoretical predictions given by \eqref{mkThe} (dashed lines) for $\epsilon=0.065, 0.072, 0.086, 0.101.$}
 \label{mkFig} 
\end{figure}

    In Figure \ref{mkFig} we compare this prediction for $\omega_k|\langle
    m_k^{(a)}(t)\rangle_t|$ in thermal equilibrium to the values given
    by numerical simulations {for varying values of nonlinearity:} the results are in good agreement in the
    weakly nonlinear regime, $\epsilon<0.1$ {Here 500 ensembles were used to build the correlator $m_k(t)$; the subsequent time averaging
    window used was $10^5$ with a sample spacing of $\Delta t=0.1$.}
    For larger nonlinearity, is expected that higher order terms play
    a role in the evolution of the anomalous correlator.

\begin{figure} [h] 
\label{fig:T}
\centering
\includegraphics[width=0.9\linewidth]{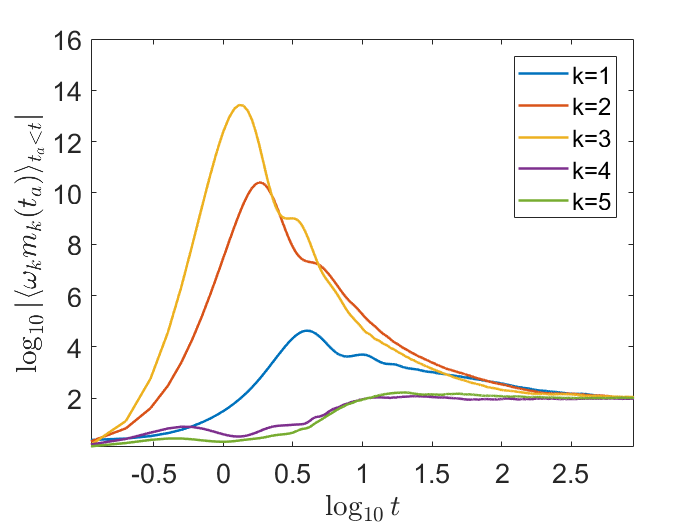}
\caption{{\color{black}{Time evolution of the first five modes of the averaged quantity 
$|\langle \omega_k m_k^{(a)}(t_a)\rangle_{t_a<t}|$  from $t=0$ to $t=1000$, with $\epsilon=10$.
}}}
 \label{mktimeev} 
\end{figure}

    In Figure \ref{mktimeev} we show the time evolution of
    the first five modes of $|\langle \omega_k m_k^{(a)}(t_a)\rangle_{t_a<t}|$, {{where ensemble averaging is used to build $m_k^{(a)}(t)$ and time averaging is used over the window $0<t_a<t$ to remove fast oscillations (see equation \eqref{mkafull}). We verify that this
    quantity is indeed initially zero due to the randomness of
    phases.}} Here we use a larger value of nonlinearity $\epsilon=10$ to show the development of the anomalous correlator in a shorter time window. The amplitudes were initialized so that
    $|a_k(t=0)|=\sqrt{\frac{N-k}{2}}$ for $k=\pm 1,\pm 2,\pm 3$, with
    higher modes zero. The phases were initially normally distributed.
    We observe that the anomalous correlator grows with time, reaching
    a peak in modes 1, 2, 3, before it eventually saturates between all
    modes equally, with $\omega_k |\langle m^{(a)}_k(t_a)\rangle_{t<t_a}|$ being
    constant for large times, {as expected from our prediction \eqref{mkEq}.}
    
\begin{figure} [h] 
\centering
\includegraphics[width=0.9\linewidth]{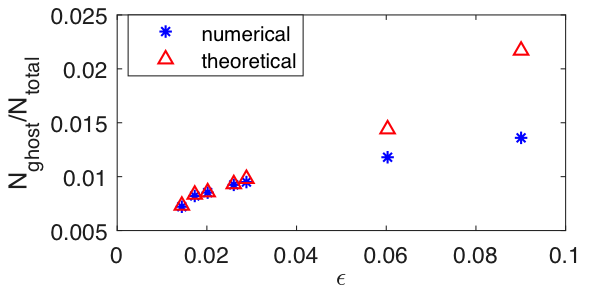}\hfil
\caption{The ratio between the number of ``ghost'' excitations,
  $\text{N}_{\text{ghost}}$, over the total number of waves,
  $\text{N}_{\text{tot}}$, as a function of the
  nonlinearity. Numerical values are blue; theoretical values, red
  triangles, are given by \eqref{percentageGhost}.}
\label{figGhostRatio}
\end{figure}

\section{Theoretical prediction for ``ghost'' excitations} 
We have now developed all the tools for predicting  analytically the 
$(k-\Omega)$ spectrum as defined in the equation (\ref{eq:kospectrum}).
Taking the Fourier Transform in time of the canonical transformation
(see appendix C), using the generalized Wick's decomposition and the hypothesis of statistical stationarity and 
homogeneity, we get at leading order:
\begin{equation}
\begin{split}
N^{(a)}(k,\Omega)=&n^{(b)}(k,t_0)\delta(\Omega-\omega_k)+\\
&+F(k) {\rm Re}[ m^{(b)}(k,t_0)]\delta(\Omega+\omega_k)
\end{split}	
\label{eq:kospectrum1}
\end{equation}
with
\begin{equation}
\begin{split}
&F(k)=4\int\sum_{l} B_{klkl}^{(3)} N^{(b)}(l,\Omega_p) d\Omega_p,
\end{split}
\label{eq:kospectrum2}
\end{equation}
where we have used the fact that at the leading order
$m^{(b)}(k,t)\simeq m^{(a)}(k,t)$ and $n^{(b)}(k,t_0)\simeq
n^{(a)}(k,t_0)$.  The equation (\ref{eq:kospectrum1})  predicts the presence of the upper and lower
branch in the $(k-\Omega$) plane. The presence of ``ghost''
excitations is clearly related to the second-order anomalous
correlator.  We can now predict the percentage of ``ghost''
excitations as
\begin{equation}
\frac{\text{N}_{\text{ghost}}}{\text{N}_{\text{tot}}}=
\frac{ \sum_k {\rm Re}[m^{(a)}(k,t_0)] F(k)}
{\sum_{k}\big(n^{(a)}(k,t_0)+  {\rm Re}[m^{(a)}(k,t_0)]F(k)\big)}
\label{percentageGhost}
\end{equation}
{In Figure \ref{figGhostRatio} we plot the ratio as determined by \eqref{percentageGhost} compared with the ratio observed in our simulations for several values of nonlinearity. We find that the results agree for small values of nonlinearity $\epsilon<0.03$; for larger nonlinearity the theoretical prediction is considerably larger.}
\section{Nonlinear standing waves}
\begin{figure}
\label{fig:T}
\centering
\includegraphics[width=1.\linewidth]{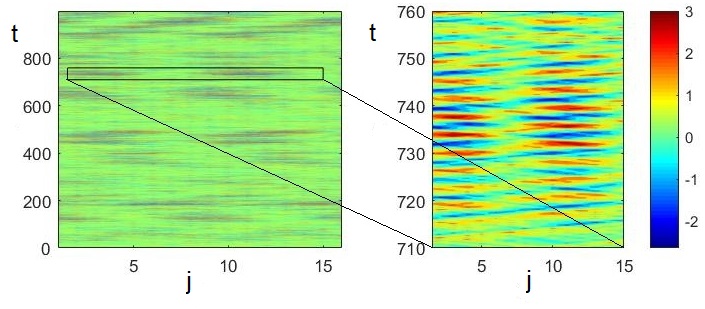}
\caption{Color map of the displacement $q_j(t)$ for the system with
  $\epsilon=4.74$ initialized with particles at rest with initial
  positions as a single sine wave. A nonlinear standing wave pattern
  is visible.}
 \label{standingWavesFig} 
\end{figure}

The development of a regime characterized by an anomalous spectrum
corresponds to a tendency for the system to develop standing waves in
the original displacement variable $q_j(t)$. Indeed, the existence of
an anomalous spectrum implies a correlation between positive and
negative wave numbers. {{ The connection between the
    anomalous correlator and standing waves can be seen on the
    following illustrative example. Consider the restrictive ensemble
    of realizations of the {\it linear} system where amplitudes
    and phases are initiated in Fourier
    space with a correlation between wave numbers $k=1$ and $k=-1$.
    Namely, let us initialize the system with  amplitudes are equal and phases
    given by
\begin{align}
    a_k(t=0)= 
\begin{cases}
    A_1 e^{-i\phi_1},& \text{if } k=1\\
    A_1 e^{i\phi_1},& \text{if } k=-1\\
    0,              & \text{otherwise},
\end{cases}
\label{simpleIC}
\end{align}
with the random phase $\phi_1$. 
In terms of the displacement variables this would correspond to the system being initially at rest and displaced from equilibrium as a single wave
\begin{align*}
    q_j(t=0)=2A_1\sqrt{\frac{2}{\omega_1}}\cos\Big(\frac{2 \pi j}{N}-\phi_1\Big).
\end{align*}

Since the system is assumed to be linear, the time evolution of
complex amplitudes $a_{1}$ and $a_{-1}$ will be given by
\begin{align*}a_{\pm 1}(t)\simeq A_1 e^{-i (\omega_1 t\pm \phi_i)}.
\end{align*}
Averaging over random phase $\phi_1$, the anomalous correlator becomes
$$m_1 = \langle a_1 a_{-1}\rangle = A_1^2 e^{-2 i \omega_1 t},$$
analogous to the oscillating term of equation \eqref{mkafull} for the anomalous correlator in the nonlinear case with amplitudes and phases being initially completely random.
In terms of displacement, such initial conditions give
\begin{align*}
    q_j(t)=2A_1\sqrt{\frac{2}{\omega_1}}\cos\Big(\frac{2 \pi j}{N}-\phi_1\Big)\cos(\omega_1 t),
\end{align*}
which corresponds to the standing wave pattern.  Thus, we see that the
phase and amplitude correlations which result in a nonzero anomalous
correlator are  directly linked to the formation of standing
waves in this particular example. 

This consideration can be generalized for the case of weakly nonlinear
systems and more general initial conditions. Indeed, for weakly
nonlinear systems the amplitudes $|a_1|$ and $|a_{-1}|$ will be
changing slowly over many oscillations, thus maintaining strongly
nonzero anomalous correlator and standing waves.}

In Figure 
\ref{standingWavesFig}-(a)
we numerically solve the equations of motion with initial conditions given by \eqref{simpleIC}.  Here we plot a colormap of the displacement $q_j(t)$ for all masses as a function of time as
the system reaches the timescale required for statistical thermal equilibrium. The nonlinearity parameter $\epsilon=4.74$, in the regime of strong nonlinearity and outside the regime of validity of our theory.  Nevertheless, we initially consider this example to display how the system behaves when the phase correlations develop rapidly. The existence of several regions
of standing wave behavior are clearly visible as darker regions in the image, as the inset Figure \ref{standingWavesFig}-(b) shows. 

It is important to emphasize that Figure \ref{standingWavesFig} shows a single realization of the system, while correlators $m_k(t), n_k(t)$ describe statistical ensemble--averaged quantities. Thus the existence of the standing wave patterns is not in violation of the presumed assumption of spatial homogeneity. 

Below we give numerical evidence that  such coherent structures can also be observed for smaller values of nonlinearity that are within the regime of validity of our theory.}

\begin{figure*}
\label{fig:T}
\centering
\includegraphics[width=1\linewidth]{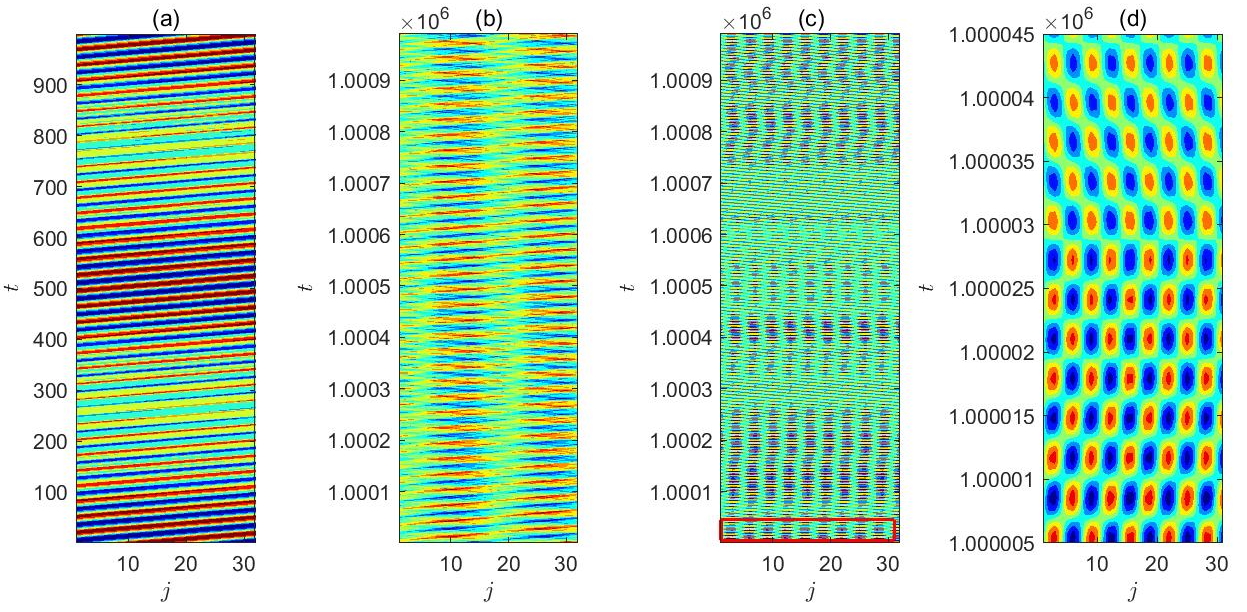}
  \caption{Color map of the displacement $q_j(t)$ for $\epsilon=0.02$: (a) initial traveling wave, $0<t<1000$ (b) standing wave structure in thermal equilibrium, $10^6<t<10^6+1000$ (c) $\tilde{q}_j(t)$, displacement after removing wave numbers $k=1...4, 7...N$ (d) closer look at the boxed region in (c).}
 \label{standWaves02} 
\end{figure*}

In Figure \ref{standWaves02}-(a) we plot the displacement as a function of time for the system with $\epsilon=0.02$, { a value of nonlinearity well within the regime of agreement of our theory as shown in Figures \ref{figGhostRatio}, \ref{mkFig}. Here, we prescribe initial conditions so that the total energy is initially in the first wave number, i.e. $a_k(t=0)=0$ for all $k\neq1$, and plot a single realization.  This corresponds to a pure traveling wave solution in the linear system; indeed as seen in Figure \ref{standWaves02}-(a), the system is initially a traveling wave, represented by series of slanted parallel lines in the colormap of $q_j(t)$. Conversely, in Figure \ref{standWaves02}-(b) we show that by the time the system has reached the timescale required for statistical thermal equilibrium, a prominent standing wave has developed, due to the phase correlations between positive and negative lowest wave numbers. Notably, phase correlations are not restricted to only the lowest wave numbers.} To emphasize {{this}}, we consider the following { spatial frequency filter} applied to the displacement 
\begin{equation}
\tilde{q}_j(t)=\sum_{k=-N/2+1}^{N/2}  H_kQ_k e^{ i 2\pi j k/N}, \label{DFT}
\end{equation}
where {$H_k=\begin{cases}
    1,& \text{if } k=5,6\\
    0,              & \text{otherwise}
\end{cases}$,  is selected to only show the waves with frequencies corresponding to $k=5,6$.}

{{We plot the resulting colormap of $\tilde{q}_j(t)$ in Figures \ref{standWaves02}-(c) and its inset \ref{standWaves02}-(d)}. Here we clearly still observe these standing waves in the selected unfiltered wave numbers, meaning that the coherent structures are not limited to the lowest wave number.  Our choice of displaying wave numbers $5,6$ is arbitrary; we also verified that similar structures exist over all the wave numbers.}

\begin{figure*}
\label{fig:T}
\centering
\includegraphics[width=1\linewidth]{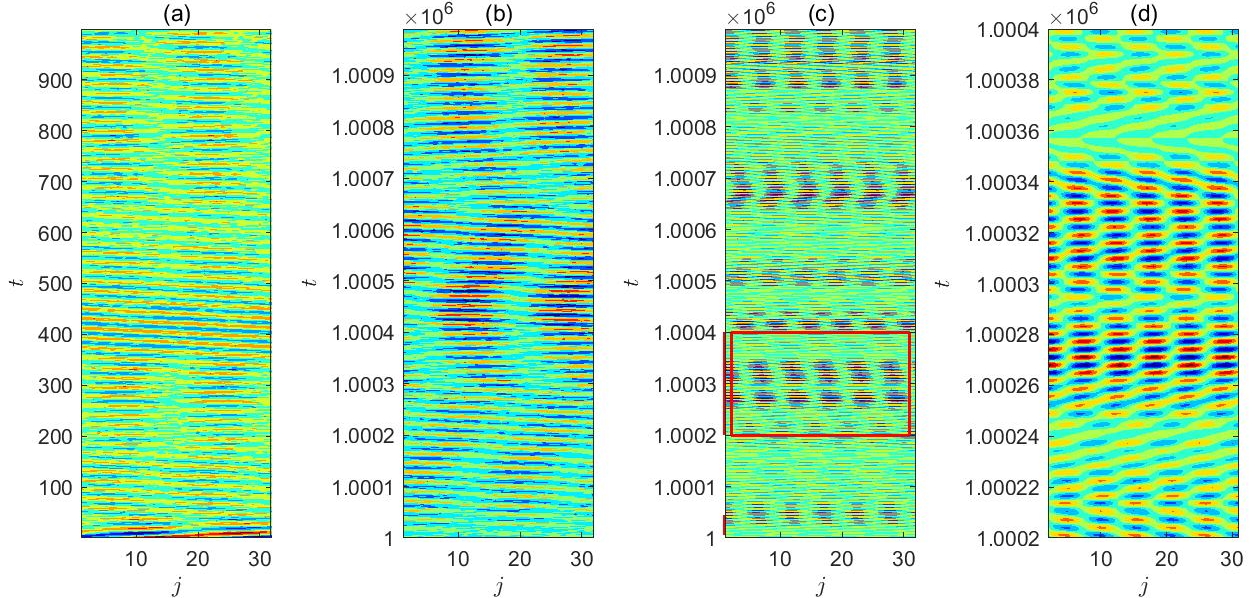}
  \caption{Color map of the displacement $q_j(t)$ for $\epsilon=0.54$: (a) initial traveling wave, $0<t<1000$ (b) standing wave structure in thermal equilibrium, $10^6<t<10^6+1000$ (c) $\tilde{q}_j(t)$, displacement after removing wave numbers $k=1...3, 5...N$ (d) closer look at the boxed region in (c).}
 \label{standWaves54} 
\end{figure*}

Similarly, in Figure \ref{standWaves54} we obtain similar results for a moderate value of nonlinearity $\epsilon=0.54$ just outside the range of applicability of our theory.  { We plot the initial time evolution of the displacements in Figure \ref{standWaves54}-(a), the time evolution of the displacements in thermal equilibrium in Figure \ref{standWaves54}-(b), and the displacements after applying a spatial frequency filter to emphasize wave number $k=4$ in Figure \ref{standWaves54}-(c). We note the arbitrary fluctuations between the coherent standing waves and between traveling waves in Figure \ref{standWaves54}-(b,c).}  

\section{Conclusion}
In this paper we have given the numerical evidence that anomalous
correlators develop spontaneously in a classical system. From a
theoretical point of view it is possible to develop a theory for
weakly nonlinear dispersive waves that accounts for presence of such
anomalous correlator. The framework in which the theory has been
developed is the Wave Turbulence one. In such theory one usually is
interested in the second order correlator $\langle a_{k_i}a_{k_j}^*
\rangle$ which is strictly related to the wave action
spectrum. However, what is clear from numerical simulations of the
$\beta$-FPUT system is that also the correlator $\langle
a_{k_i}a_{k_j} \rangle$ can assume values that are different from
zero. This finding has consequences on the standard Wave Turbulence
theory that is based on the Wick's selection rule, i.e. the splitting
of higher order correlators as a sum of products of second order
correlators. { Following \cite{falkovich1992kolmogorov,
    LVOVBOOK}, we have generalized the Wick's rule by including the
  anomalous correlators.  We note that we differ from the case
  described in the S--theory \cite{zakharov1975spin,l2012wave} in that
  there the existence of anomalous correlators was connected with
  coherent pumping in the system, with the anomalous correlator being
  a measure of partial coherence for exiting waves. In our
  observations and predictions, waves with random initial conditions
  form phase correlations with each other, resulting in an anomalous
  correlator which is initially zero but then saturates to a nonzero
  value as it evolves with time.}

One of the most striking
manifestation of those correlators is the appearance of ``ghost
excitations'', i.e. those characterized by a negative frequencies. A
formula for the content of energy of such excitations as a function of
the wave spectrum is obtained and compared favorably, in the weakly
nonlinear, regime with numerical simulations. Moreover, we have shown that the spontaneous emergence of the anomalous correlator is strongly connected with the formation of nonlinear standing waves; indeed, the presence of those waves implies a strong correlation between the phases of positive and negative wave numbers.

Our approach paves a new road to investigate dispersive nonlinear
systems by taking into account not only amplitudes of the waves, as in
traditional wave turbulence, but also the phases of the waves. {We
conjecture that the anomalous correlators play an important role in
the theory of extreme events, such as rogue waves, which form via a mechanism related to phase locking between different wave numbers \cite{Onorato2013a}. Phase locking also leads  to the existence of solitons in nonlinear media.

As was discussed in the introduction, anomalous phase correlations have been observed to play a role in causing shifts of wave action from turbulence and condensate in the Nonlinear Schrodinger equation \cite{NLSPhaseCorr}.  Our approach of extending Wave Turbulence Theory to include the anomalous correlator could be generalized to address the role these correlations play in the statistical properties of the Nonlinear Schrodinger equation and other integrable systems. On a similar note, recurrences in an NLS-like model were shown to be directly related to the formation of anomalous phase correlations \cite{guasoni2017incoherent}; further investigating FPUT recurrences potential ties to the anomalous correlator is a subject of current work.  

Finally, we emphasize that the Hamiltonian we considered is of the same}
family as the one for surface gravity waves (after removing by a
canonical transformation nonresonant three wave interactions). We
predict that also the anomalous correlators will play an important role
in the understanding of statistical properties of ocean waves.

{\bf Acknowledgments} { The authors are grateful to
  Dr. B. Giulinico for discussions. We are grateful to 
  anonymous referees, who's insightful suggestions improved the manuscript
  considerably.} M. O. has been funded by Progetto di Ricerca
d'Ateneo CSTO160004, by the ``Departments of Excellence 2018/2022''
Grant awarded by the Italian Ministry of Education, University and
Research (MIUR) (L.232/2016) and by Simons Foundation, Wave
Turbulence. JZ and YL acknowledge support from NSF OCE grant
1635866. YL acknowledges support from ONR grant N00014-17-1-2852.

{\bf Appendix A}\\
{\it Matrix element in eq.  (\ref{eq:fourwaveint}) - }
The matrix elements governing four wave interactions for the variable $a_k(t)$ are:
\begin{equation}
\begin{split}
&T^{(1)}_{1234}=-\frac{1}{4}\beta{\mathrm{e}^{i \pi (-k_1+k_2+k_3+k_4)/N}} \prod_{i=1}^4 \frac{2 \sin(\pi k_i/N)}{\sqrt{\omega_i}},\\
&T^{(2)}_{1234}=-3 T^{(1)}_{1-234},\;\; T^{(3)}_{1234}=3T^{(1)}_{4231}\;\;T^{(4)}_{1234}=-T^{(1)}_{-1234}.
\end{split}
\label{matrixelements}
\end{equation}

{\it Matrix elements in the canonical transformation, eq.  (\ref{CanonicalTrans})  - }
The coefficients in eq. (\ref{eq:fourwaveint}) suitable for removing nonresonant terms are given 
by:
\begin{equation}
\begin{split}
    B^{(1)}_{1234}=\frac{T^{(1)}_{1234}}{\omega_1-\omega_2-\omega_3-\omega_4},\\
    B^{(3)}_{1234}=\frac{T^{(3)}_{1234}}{\omega_4-\omega_1-\omega_2-\omega_3},\\
    B^{(4)}_{1234}=\frac{T^{(4)}_{1234}}{-\omega_1-\omega_2-\omega_3-\omega_4}.
\end{split}
\end{equation}

{\bf Appendix B}

Starting from the transformation in (\ref{CanonicalTrans})
and the generalized Wick's decomposition in (\ref{eq:wick}), we obtain:
\begin{equation}
\begin{split}
&m_k^{(a)}(t)=m_k^{(b)}(t)+ 
2  \left(n_k^{(b)}(t)+
n_{-k}^{(b)}(t)\right) \\
&\times \sum_{j} B^{(3)}_{k,{{-}k},j,j}n_{j}^{(b)}(t),
\label{can_m}
\end{split}
\end{equation}
where higher order terms in $m_k$ have been neglected.  The next step
consists in building the evolution equation for $m_k^{(b)}(t)$ from
equation (\ref{eq:zakh}). Interestingly, the evolution equation for
$m_k^{(b)}(t)$ appears as a deterministic dispersive non homogeneous
wave evolution equation \cite{LVOVBOOK},
\begin{equation}
\begin{split}
&i \frac{ d m_k^{(b)}} {d t}=2 \tilde\omega_k m_k^{(b)}+\big[
(n_k^{(b)}+n_{-k}^{(b)}) \sum_j T_{k-kj-j}m_j^{(b)}\big],
\end{split}
\label{mkdot}
\end{equation}
with 
 $ \tilde\omega_k=\omega_k+ 2 \sum_j T_{kjkj}n_j^{(b))}$. Such equations have been derived in the theory of Bose Einstein condensates and superconductivity.

The equation for the spectrum, see \cite{LVOVBOOK}, is given by
\begin{equation}
\begin{split}
& \frac{ d n_k^{(b)}} {d t}= -2 {\rm Im }[m_k^{(b)} \sum T_{k,-k,j,-j} m_j^{(b)*}].
\end{split}
\label{nkdot}
\end{equation}
From equations (\ref{mkdot}) and (\ref{nkdot}), after some algebra, it is possible to 
show that the following interesting relations holds:
\begin{equation}
\begin{split}
&\frac{d [|m^{(b)}(k)|^2]}{d{t}}=\frac{d [n_k^{(b)}n_{-k}^{(b)}] }{d{t}}\\
&\frac{d [n_k^{(b)}-n_{-k}^{(b)}]}{d{t}}=0
\end{split}
\end{equation}
If $n_k^{(b)}$ has reached energy equipartition such that 
$n_k^{(b)}=n_{-k}^{(b)}=const/\omega_k$ then 
$|m_k^{(b)}|=n_k^{(b)}$; therefore, we expect to observe equipartition also for
$\omega_k |m_k^{(b)}|$.

We now consider the leading order solution of equation (\ref{mkdot})
\begin{equation}
m_k^{(b)}(t)=m_k^{(b)}(t_0) e^{-i 2\tilde\omega_k t}+ \text{higher order
  terms}.  \nonumber
\end{equation}
and plug it in (\ref{can_m}), and assuming that the spectrum $n_k$ is in 
stationary conditions, we get
\begin{equation}
\begin{split}
 & m_k^{(a)}(t)=m_k^{(b)}(t_0) e^{-i 2\tilde \omega_k t} + 2  \left(n_k^{(a)}(t_0)+n_{-k}^{(a)}(t_0)\right)\\
 &
\times \sum_{j} B^{(3)}_{k,{{-}k},j,j}n_{j}^{(a)}(t_0).
\end{split}
\label{mkafull}
\end{equation}
Note that we have used the fact that at the leading order $n_k^{(b)}(t_0)\simeq n_k^{(a)}(t_0)$.
\\
{\bf Appendix C}\\
We consider equation eq. (\ref{CanonicalTrans}) and take the Fourier Transform in time, to get:
\begin{equation}
\begin{split}
 &a_{k_i,\Omega_p}=b_{k_i,\Omega_p}+\\
 &+\int \sum_{j, k,l } B_{ijkl}^{(1)}b_{{j},\Omega_q}b_{{k},{\Omega_r}}b_{{l},{\Omega_s}}\delta_{ij}^{kl} 
 \delta_{\Omega_p}^{\Omega_q\Omega_r\Omega_s}
 d\Omega_{qrs}+
 \\
 &+\int \sum_{j, k,l } B_{ijkl}^{(3)}b_{{j},\Omega_q}^*b_{{k},{\Omega_r}}^*b_{{l},{\Omega_s}}\delta_{ijk}^{l} 
  \delta_{\Omega_p\Omega_q}^{\Omega_r\Omega_s}
 d\Omega_{qrs}+
\\
 &+\int \sum_{j, k,l } B_{ijkl}^{(4)}b_{{j},\Omega_q}^*b_{{k},{\Omega_r}}^*b_{{l},{\Omega_s}}^*\delta_{ijkl} 
  \delta_{\Omega_p\Omega_q\Omega_r\Omega_s}
 d\Omega_{qrs}.
 \end{split}
  \end{equation}
The next step is to build the second order correlator 
 $\langle   a(k_i,\Omega_l)  a(k_j,\Omega_m)^*\rangle$ assuming stationarity. 

We use the generalized Wick's decomposition in (\ref{eq:wick}), i.e. including the anomalous correlators. The leading order result is contained in equation (\ref{eq:kospectrum1}).
\clearpage


\begin{thebibliography}{26}%
\makeatletter
\providecommand \@ifxundefined [1]{%
 \@ifx{#1\undefined}
}%
\providecommand \@ifnum [1]{%
 \ifnum #1\expandafter \@firstoftwo
 \else \expandafter \@secondoftwo
 \fi 
}%
\providecommand \@ifx [1]{%
 \ifx #1\expandafter \@firstoftwo
 \else \expandafter \@secondoftwo
 \fi
}%
\providecommand \natexlab [1]{#1}%
\providecommand \enquote  [1]{``#1''}%
\providecommand \bibnamefont  [1]{#1}%
\providecommand \bibfnamefont [1]{#1}%
\providecommand \citenamefont [1]{#1}%
\providecommand \href@noop [0]{\@secondoftwo}%
\providecommand \href [0]{\begingroup \@sanitize@url \@href}%
\providecommand \@href[1]{\@@startlink{#1}\@@href}%
\providecommand \@@href[1]{\endgroup#1\@@endlink}%
\providecommand \@sanitize@url [0]{\catcode `\\12\catcode `\$12\catcode
  `\&12\catcode `\#12\catcode `\^12\catcode `\_12\catcode `\%12\relax}%
\providecommand \@@startlink[1]{}%
\providecommand \@@endlink[0]{}%
\providecommand \url  [0]{\begingroup\@sanitize@url \@url }%
\providecommand \@url [1]{\endgroup\@href {#1}{\urlprefix }}%
\providecommand \urlprefix  [0]{URL }%
\providecommand \Eprint [0]{\href }%
\providecommand \doibase [0]{http://dx.doi.org/}%
\providecommand \selectlanguage [0]{\@gobble}%
\providecommand \bibinfo  [0]{\@secondoftwo}%
\providecommand \bibfield  [0]{\@secondoftwo}%
\providecommand \translation [1]{[#1]}%
\providecommand \BibitemOpen [0]{}%
\providecommand \bibitemStop [0]{}%
\providecommand \bibitemNoStop [0]{.\EOS\space}%
\providecommand \EOS [0]{\spacefactor3000\relax}%
\providecommand \BibitemShut  [1]{\csname bibitem#1\endcsname}%
\let\auto@bib@innerbib\@empty
\bibitem [{\citenamefont {Falkovich}\ \emph {et~al.}(1992)\citenamefont
  {Falkovich}, \citenamefont {Lvov},\ and\ \citenamefont
  {Zakharov}}]{falkovich1992kolmogorov}%
  \BibitemOpen
  \bibfield  {author} {\bibinfo {author} {\bibfnamefont {G.}~\bibnamefont
  {Falkovich}}, \bibinfo {author} {\bibfnamefont {V.~S.}\ \bibnamefont {Lvov}},
  \ and\ \bibinfo {author} {\bibfnamefont {V.~E.}\ \bibnamefont {Zakharov}},\
  }\href@noop {} {\emph {\bibinfo {title} {Kolmogorov spectra of turbulence}}}\
  (\bibinfo  {publisher} {Springer, Berlin},\ \bibinfo {year}
  {1992})\BibitemShut {NoStop}%
\bibitem [{\citenamefont {Nazarenko}(2011)}]{nazarenko2011wave}%
  \BibitemOpen
  \bibfield  {author} {\bibinfo {author} {\bibfnamefont {S.}~\bibnamefont
  {Nazarenko}},\ }\href@noop {} {\emph {\bibinfo {title} {Wave turbulence}}},\
  Vol.\ \bibinfo {volume} {825}\ (\bibinfo  {publisher} {Springer},\ \bibinfo
  {year} {2011})\BibitemShut {NoStop}%
\bibitem [{\citenamefont {Benney}\ and\ \citenamefont {Newell}(1969)}]{Newell}%
  \BibitemOpen
  \bibfield  {author} {\bibinfo {author} {\bibfnamefont {J.}~\bibnamefont
  {Benney}}\ and\ \bibinfo {author} {\bibfnamefont {A.~C.}\ \bibnamefont
  {Newell}},\ }\bibfield  {title} {\enquote {\bibinfo {title} {Random wave
  closure},}\ }\href@noop {} {\bibfield  {journal} {\bibinfo  {journal}
  {Studies in Appl. Math.}\ }\textbf {\bibinfo {volume} {48}},\ \bibinfo
  {pages} {1} (\bibinfo {year} {1969})}\BibitemShut {NoStop}%
\bibitem [{\citenamefont {Newell}(1968)}]{N1}%
  \BibitemOpen
  \bibfield  {author} {\bibinfo {author} {\bibfnamefont {A.~C.}\ \bibnamefont
  {Newell}},\ }\bibfield  {title} {\enquote {\bibinfo {title} {The closure
  problem in a system of random gravity waves},}\ }\href@noop {} {\bibfield
  {journal} {\bibinfo  {journal} {Review of Geophysics}\ }\textbf {\bibinfo
  {volume} {6}},\ \bibinfo {pages} {1} (\bibinfo {year} {1968})}\BibitemShut
  {NoStop}%
\bibitem [{\citenamefont {Benney}\ and\ \citenamefont {Saffmann}(1966)}]{Ben}%
  \BibitemOpen
  \bibfield  {author} {\bibinfo {author} {\bibfnamefont {D.~J.}\ \bibnamefont
  {Benney}}\ and\ \bibinfo {author} {\bibfnamefont {P.}~\bibnamefont
  {Saffmann}},\ }\bibfield  {title} {\enquote {\bibinfo {title} {Nonlinear
  interaction of random waves in a dispersive medium},}\ }\href@noop {}
  {\bibfield  {journal} {\bibinfo  {journal} {Proc Royal. Soc}\ }\textbf
  {\bibinfo {volume} {289}},\ \bibinfo {pages} {301--320} (\bibinfo {year}
  {1966})}\BibitemShut {NoStop}%
\bibitem [{\citenamefont {Kadomtsev}(1965)}]{K}%
  \BibitemOpen
  \bibfield  {author} {\bibinfo {author} {\bibfnamefont {B.~B.}\ \bibnamefont
  {Kadomtsev}},\ }\href@noop {} {\emph {\bibinfo {title} {Plasma Turbulence}}}\
  (\bibinfo  {publisher} {Academic Press},\ \bibinfo {address} {New York},\
  \bibinfo {year} {1965})\BibitemShut {NoStop}%
\bibitem [{\citenamefont {Choi}\ \emph {et~al.}(2005)\citenamefont {Choi},
  \citenamefont {Lvov}, \citenamefont {Nazarenko},\ and\ \citenamefont
  {Pokorni}}]{X1}%
  \BibitemOpen
  \bibfield  {author} {\bibinfo {author} {\bibfnamefont {Y.}~\bibnamefont
  {Choi}}, \bibinfo {author} {\bibfnamefont {Y.~V.}\ \bibnamefont {Lvov}},
  \bibinfo {author} {\bibfnamefont {S.}~\bibnamefont {Nazarenko}}, \ and\
  \bibinfo {author} {\bibfnamefont {B.}~\bibnamefont {Pokorni}},\ }\bibfield
  {title} {\enquote {\bibinfo {title} {Anomalous probability of large
  amplitudes in wave turbulence},}\ }\href@noop {} {\bibfield  {journal}
  {\bibinfo  {journal} {Physics Letters A}\ }\textbf {\bibinfo {volume}
  {339}},\ \bibinfo {pages} {361} (\bibinfo {year} {2005})}\BibitemShut
  {NoStop}%
\bibitem [{\citenamefont {Choi}\ \emph {et~al.}(2004)\citenamefont {Choi},
  \citenamefont {Lvov},\ and\ \citenamefont {Nazarenko}}]{X2}%
  \BibitemOpen
  \bibfield  {author} {\bibinfo {author} {\bibfnamefont {Y.}~\bibnamefont
  {Choi}}, \bibinfo {author} {\bibfnamefont {Y.~V.}\ \bibnamefont {Lvov}}, \
  and\ \bibinfo {author} {\bibfnamefont {S.}~\bibnamefont {Nazarenko}},\
  }\bibfield  {title} {\enquote {\bibinfo {title} {Probability densities and
  preservation of randomness in wave turbulence},}\ }\href@noop {} {\bibfield
  {journal} {\bibinfo  {journal} {Physics Letters A}\ }\textbf {\bibinfo
  {volume} {332}},\ \bibinfo {pages} {230} (\bibinfo {year}
  {2004})}\BibitemShut {NoStop}%
\bibitem [{\citenamefont {Lvov}(2012)}]{l2012wave}%
  \BibitemOpen
  \bibfield  {author} {\bibinfo {author} {\bibfnamefont {V.~S.}\ \bibnamefont
  {Lvov}},\ }\href@noop {} {\emph {\bibinfo {title} {Wave turbulence under
  parametric excitation: applications to magnets}}}\ (\bibinfo  {publisher}
  {Springer Science \& Business Media},\ \bibinfo {year} {2012})\BibitemShut
  {NoStop}%
\bibitem [{\citenamefont {Zakharov}\ \emph {et~al.}(1975)\citenamefont
  {Zakharov}, \citenamefont {Lvov},\ and\ \citenamefont
  {Starobinets}}]{zakharov1975spin}%
  \BibitemOpen
  \bibfield  {author} {\bibinfo {author} {\bibfnamefont {V.~E.}\ \bibnamefont
  {Zakharov}}, \bibinfo {author} {\bibfnamefont {V.~S.}\ \bibnamefont {Lvov}},
  \ and\ \bibinfo {author} {\bibfnamefont {S.~S.}\ \bibnamefont
  {Starobinets}},\ }\bibfield  {title} {\enquote {\bibinfo {title} {Spin-wave
  turbulence beyond the parametric excitation threshold},}\ }\href@noop {}
  {\bibfield  {journal} {\bibinfo  {journal} {Physics-Uspekhi}\ }\textbf
  {\bibinfo {volume} {17}},\ \bibinfo {pages} {896--919} (\bibinfo {year}
  {1975})}\BibitemShut {NoStop}%
\bibitem [{\citenamefont {Janssen}(2004)}]{janssenbook04}%
  \BibitemOpen
  \bibfield  {author} {\bibinfo {author} {\bibfnamefont {P.}~\bibnamefont
  {Janssen}},\ }\href@noop {} {\emph {\bibinfo {title} {{The interaction of
  ocean waves and wind}}}}\ (\bibinfo  {publisher} {Cambridge University
  Press},\ \bibinfo {address} {Cambridge},\ \bibinfo {year} {2004})\ p.\
  \bibinfo {pages} {379}\BibitemShut {NoStop}%
\bibitem [{\citenamefont {Picozzi}\ \emph {et~al.}(2014)\citenamefont
  {Picozzi}, \citenamefont {Garnier}, \citenamefont {Hansson}, \citenamefont
  {Suret}, \citenamefont {Randoux}, \citenamefont {Millot},\ and\ \citenamefont
  {Christodoulides}}]{Picozzi2014}%
  \BibitemOpen
  \bibfield  {author} {\bibinfo {author} {\bibfnamefont {A.}~\bibnamefont
  {Picozzi}}, \bibinfo {author} {\bibfnamefont {J.}~\bibnamefont {Garnier}},
  \bibinfo {author} {\bibfnamefont {T.}~\bibnamefont {Hansson}}, \bibinfo
  {author} {\bibfnamefont {P.}~\bibnamefont {Suret}}, \bibinfo {author}
  {\bibfnamefont {S.}~\bibnamefont {Randoux}}, \bibinfo {author} {\bibfnamefont
  {G.}~\bibnamefont {Millot}}, \ and\ \bibinfo {author} {\bibfnamefont {D.~N.}\
  \bibnamefont {Christodoulides}},\ }\bibfield  {title} {\enquote {\bibinfo
  {title} {Optical wave turbulence: Towards a unified nonequilibrium
  thermodynamic formulation of statistical nonlinear optics},}\ }\href@noop {}
  {\bibfield  {journal} {\bibinfo  {journal} {Physics Reports}\ }\textbf
  {\bibinfo {volume} {542}} (\bibinfo {year} {2014})}\BibitemShut {NoStop}%
\bibitem [{\citenamefont {Proment}\ \emph {et~al.}(2012)\citenamefont
  {Proment}, \citenamefont {Nazarenko},\ and\ \citenamefont
  {Onorato}}]{Proment2012}%
  \BibitemOpen
  \bibfield  {author} {\bibinfo {author} {\bibfnamefont {D.}~\bibnamefont
  {Proment}}, \bibinfo {author} {\bibfnamefont {S.}~\bibnamefont {Nazarenko}},
  \ and\ \bibinfo {author} {\bibfnamefont {M.}~\bibnamefont {Onorato}},\
  }\bibfield  {title} {\enquote {\bibinfo {title} {Sustained turbulence in the
  three-dimensional grosspitaevskii model},}\ }\href {\doibase
  10.1016/j.physd.2011.06.007} {\bibfield  {journal} {\bibinfo  {journal}
  {Physica D: Nonlinear Phenomena}\ }\textbf {\bibinfo {volume} {241}}
  (\bibinfo {year} {2012}),\ 10.1016/j.physd.2011.06.007}\BibitemShut {NoStop}%
\bibitem [{\citenamefont {M.~Onorato}\ and\ \citenamefont {Lvov}(2015)}]{PNAS}%
  \BibitemOpen
  \bibfield  {author} {\bibinfo {author} {\bibfnamefont {D.~Proment}\
  \bibnamefont {M.~Onorato}, \bibfnamefont {L.~Vozella}}\ and\ \bibinfo
  {author} {\bibfnamefont {Y.~V.}\ \bibnamefont {Lvov}},\ }\bibfield  {title}
  {\enquote {\bibinfo {title} {A route to thermalization in the
  $\alpha$-fermi-pasta-ulam system},}\ }\href@noop {} {\bibfield  {journal}
  {\bibinfo  {journal} {Proceeding of National Academy of Science}\ }\textbf
  {\bibinfo {volume} {112}},\ \bibinfo {pages} {4208--4213} (\bibinfo {year}
  {2015})}\BibitemShut {NoStop}%
\bibitem [{\citenamefont {V.S.Lvov}(1994)}]{LVOVBOOK}%
  \BibitemOpen
  \bibfield  {author} {\bibinfo {author} {\bibnamefont {V.S.Lvov}},\
  }\href@noop {} {\emph {\bibinfo {title} {Wave Turbulence Under Parametric
  Excitations, Applications to Magnets}}}\ (\bibinfo  {publisher}
  {Springer-Verlag},\ \bibinfo {year} {1994})\BibitemShut {NoStop}%
\bibitem [{\citenamefont {Fermi}\ \emph {et~al.}(1955)\citenamefont {Fermi},
  \citenamefont {Pasta},\ and\ \citenamefont {Ulam}}]{fermi1955studies}%
  \BibitemOpen
  \bibfield  {author} {\bibinfo {author} {\bibfnamefont {E.}~\bibnamefont
  {Fermi}}, \bibinfo {author} {\bibfnamefont {J.}~\bibnamefont {Pasta}}, \ and\
  \bibinfo {author} {\bibfnamefont {S.}~\bibnamefont {Ulam}},\ }\href@noop {}
  {\emph {\bibinfo {title} {Studies of nonlinear problems}}},\ \bibinfo {type}
  {Tech. Rep.}\ (\bibinfo  {institution} {I, Los Alamos Scientific Laboratory
  Report No. LA-1940},\ \bibinfo {year} {1955})\BibitemShut {NoStop}%
\bibitem [{\citenamefont {Pistone}\ \emph {et~al.}(2018)\citenamefont
  {Pistone}, \citenamefont {Onorato},\ and\ \citenamefont
  {Chibbaro}}]{pistone2018thermalization}%
  \BibitemOpen
  \bibfield  {author} {\bibinfo {author} {\bibfnamefont {L.}~\bibnamefont
  {Pistone}}, \bibinfo {author} {\bibfnamefont {M.}~\bibnamefont {Onorato}}, \
  and\ \bibinfo {author} {\bibfnamefont {S.}~\bibnamefont {Chibbaro}},\
  }\bibfield  {title} {\enquote {\bibinfo {title} {Thermalization in the
  discrete nonlinear klein-gordon chain in the wave-turbulence framework},}\
  }\href@noop {} {\bibfield  {journal} {\bibinfo  {journal} {EPL (Europhysics
  Letters)}\ }\textbf {\bibinfo {volume} {121}},\ \bibinfo {pages} {44003}
  (\bibinfo {year} {2018})}\BibitemShut {NoStop}%
\bibitem [{\citenamefont {Lvov}\ and\ \citenamefont
  {Onorato}(2018)}]{lvov2018double}%
  \BibitemOpen
  \bibfield  {author} {\bibinfo {author} {\bibfnamefont {Y.~V.}\ \bibnamefont
  {Lvov}}\ and\ \bibinfo {author} {\bibfnamefont {M.}~\bibnamefont {Onorato}},\
  }\bibfield  {title} {\enquote {\bibinfo {title} {Double scaling in the
  relaxation time in the $\beta$-fermi-pasta-ulam-tsingou model},}\ }\href@noop
  {} {\bibfield  {journal} {\bibinfo  {journal} {Physical review letters}\
  }\textbf {\bibinfo {volume} {120}},\ \bibinfo {pages} {144301} (\bibinfo
  {year} {2018})}\BibitemShut {NoStop}%
\bibitem [{\citenamefont {Lee}\ \emph {et~al.}(2013)\citenamefont {Lee},
  \citenamefont {Kovacic},\ and\ \citenamefont {Cai}}]{kovacicPNAS}%
  \BibitemOpen
  \bibfield  {author} {\bibinfo {author} {\bibfnamefont {W.}~\bibnamefont
  {Lee}}, \bibinfo {author} {\bibfnamefont {G.}~\bibnamefont {Kovacic}}, \ and\
  \bibinfo {author} {\bibfnamefont {D.}~\bibnamefont {Cai}},\ }\bibfield
  {title} {\enquote {\bibinfo {title} {Generation of dispersion in
  nondispersive nonlinear waves in thermal equilibrium},}\ }\href@noop {}
  {\bibfield  {journal} {\bibinfo  {journal} {Proceedings of the National
  Academy of Sciences of the United States of America}\ }\textbf {\bibinfo
  {volume} {110}},\ \bibinfo {pages} {3237--3241} (\bibinfo {year}
  {2013})}\BibitemShut {NoStop}%
\bibitem [{\citenamefont {Zakharov}(1968)}]{zakharov68}%
  \BibitemOpen
  \bibfield  {author} {\bibinfo {author} {\bibfnamefont {V}~\bibnamefont
  {Zakharov}},\ }\bibfield  {title} {\enquote {\bibinfo {title} {{Stability of
  period waves of finite amplitude on surface of a deep fluid}},}\ }\href@noop
  {} {\bibfield  {journal} {\bibinfo  {journal} {J. Appl. Mech. Tech. Phys.}\
  }\textbf {\bibinfo {volume} {9}},\ \bibinfo {pages} {190--194} (\bibinfo
  {year} {1968})}\BibitemShut {NoStop}%
\bibitem [{\citenamefont {Bustamante}\ \emph {et~al.}(2019)\citenamefont
  {Bustamante}, \citenamefont {Hutchinson}, \citenamefont {Lvov},\ and\
  \citenamefont {Onorato}}]{bustamante2019exact}%
  \BibitemOpen
  \bibfield  {author} {\bibinfo {author} {\bibfnamefont {Miguel~D}\
  \bibnamefont {Bustamante}}, \bibinfo {author} {\bibfnamefont {Kevin}\
  \bibnamefont {Hutchinson}}, \bibinfo {author} {\bibfnamefont {Yuri~V}\
  \bibnamefont {Lvov}}, \ and\ \bibinfo {author} {\bibfnamefont {Miguel}\
  \bibnamefont {Onorato}},\ }\bibfield  {title} {\enquote {\bibinfo {title}
  {Exact discrete resonances in the fermi-pasta-ulam--tsingou system},}\
  }\href@noop {} {\bibfield  {journal} {\bibinfo  {journal} {Communications in
  Nonlinear Science and Numerical Simulation}\ }\textbf {\bibinfo {volume}
  {73}},\ \bibinfo {pages} {437--471} (\bibinfo {year} {2019})}\BibitemShut
  {NoStop}%
\bibitem [{\citenamefont {Yoshida}(1990)}]{yoshida1990construction}%
  \BibitemOpen
  \bibfield  {author} {\bibinfo {author} {\bibfnamefont {H.}~\bibnamefont
  {Yoshida}},\ }\bibfield  {title} {\enquote {\bibinfo {title} {{Construction
  of higher order symplectic integrators}},}\ }\href@noop {} {\bibfield
  {journal} {\bibinfo  {journal} {Physics Letters A}\ }\textbf {\bibinfo
  {volume} {150}},\ \bibinfo {pages} {262--268} (\bibinfo {year}
  {1990})}\BibitemShut {NoStop}%
\bibitem [{\citenamefont {Guasoni}\ \emph {et~al.}(2017)\citenamefont
  {Guasoni}, \citenamefont {Garnier}, \citenamefont {Rumpf}, \citenamefont
  {Sugny}, \citenamefont {Fatome}, \citenamefont {Amrani}, \citenamefont
  {Millot},\ and\ \citenamefont {Picozzi}}]{guasoni2017incoherent}%
  \BibitemOpen
  \bibfield  {author} {\bibinfo {author} {\bibfnamefont {M.}~\bibnamefont
  {Guasoni}}, \bibinfo {author} {\bibfnamefont {J.}~\bibnamefont {Garnier}},
  \bibinfo {author} {\bibfnamefont {B.}~\bibnamefont {Rumpf}}, \bibinfo
  {author} {\bibfnamefont {D.}~\bibnamefont {Sugny}}, \bibinfo {author}
  {\bibfnamefont {J.}~\bibnamefont {Fatome}}, \bibinfo {author} {\bibfnamefont
  {F.}~\bibnamefont {Amrani}}, \bibinfo {author} {\bibfnamefont
  {G.}~\bibnamefont {Millot}}, \ and\ \bibinfo {author} {\bibfnamefont
  {A.}~\bibnamefont {Picozzi}},\ }\bibfield  {title} {\enquote {\bibinfo
  {title} {Incoherent fermi-pasta-ulam recurrences and unconstrained
  thermalization mediated by strong phase correlations},}\ }\href@noop {}
  {\bibfield  {journal} {\bibinfo  {journal} {Physical Review X}\ }\textbf
  {\bibinfo {volume} {7}},\ \bibinfo {pages} {011025} (\bibinfo {year}
  {2017})}\BibitemShut {NoStop}%
\bibitem [{\citenamefont {Krasitskii}(1994)}]{krasitskii1994reduced}%
  \BibitemOpen
  \bibfield  {author} {\bibinfo {author} {\bibfnamefont {V~P}\ \bibnamefont
  {Krasitskii}},\ }\bibfield  {title} {\enquote {\bibinfo {title} {{On reduced
  equations in the Hamiltonian theory of weakly nonlinear surface waves}},}\
  }\href@noop {} {\bibfield  {journal} {\bibinfo  {journal} {J. Fluid Mech.}\
  }\textbf {\bibinfo {volume} {272}},\ \bibinfo {pages} {1--20} (\bibinfo
  {year} {1994})}\BibitemShut {NoStop}%
\bibitem [{\citenamefont {Onorato}\ \emph {et~al.}(2013)\citenamefont
  {Onorato}, \citenamefont {Residori}, \citenamefont {Bortolozzo},
  \citenamefont {Montina},\ and\ \citenamefont {Arecchi}}]{Onorato2013a}%
  \BibitemOpen
  \bibfield  {author} {\bibinfo {author} {\bibfnamefont {M.}~\bibnamefont
  {Onorato}}, \bibinfo {author} {\bibfnamefont {S.}~\bibnamefont {Residori}},
  \bibinfo {author} {\bibfnamefont {U.}~\bibnamefont {Bortolozzo}}, \bibinfo
  {author} {\bibfnamefont {A.}~\bibnamefont {Montina}}, \ and\ \bibinfo
  {author} {\bibfnamefont {F.~T.}\ \bibnamefont {Arecchi}},\ }\bibfield
  {title} {\enquote {\bibinfo {title} {{Rogue waves and their generating
  mechanisms in different physical contexts}},}\ }\href {\doibase
  10.1016/j.physrep.2013.03.001} {\bibfield  {journal} {\bibinfo  {journal}
  {Physics Reports}\ }\textbf {\bibinfo {volume} {528}} (\bibinfo {year}
  {2013}),\ 10.1016/j.physrep.2013.03.001}\BibitemShut {NoStop}%
  \bibitem [{\citenamefont {Bardeen}(1957)}]{BCSTheory}%
  \BibitemOpen
  \bibfield  {author} {\bibinfo {author} {\bibfnamefont {J.}~\bibnamefont
  {Bardeen}}, \bibfield  {author} {\bibinfo {author} {\bibfnamefont {L. N.}~\bibnamefont
  {Cooper}}}, \bibfield  {author} {\bibinfo {author} {\bibfnamefont {J. R.}~\bibnamefont
  {Schrieffer}}}\ }\bibfield  {title} {\enquote {\bibinfo {title} {{Microscopic Theory of Superconductivity}},}\ }\href@noop
  {} {\bibfield  {journal} {\bibinfo  {journal} {Physical Review}\
  }\textbf {\bibinfo {volume} {106}},\ \bibinfo {pages} {162--164} (\bibinfo
  {year} {1957})}\BibitemShut {NoStop}%
  \bibitem [{\citenamefont {Bardeen}(1957)}]{NLSPhaseCorr}%
  \BibitemOpen
  \bibfield  {author} {\bibinfo {author} {\bibfnamefont {P.}~\bibnamefont
  {Miller}}, \bibfield  {author} {\bibinfo {author} {\bibfnamefont {N.}~\bibnamefont
  {Vladimirova}}}, \bibfield  {author} {\bibinfo {author} {\bibfnamefont {F.}~\bibnamefont
  {Falkovich}}}\ }\bibfield  {title} {\enquote {\bibinfo {title} {{Oscillations in a turbulence-condensate system}},}\ }\href@noop
  {} {\bibfield  {journal} {\bibinfo  {journal} {Physical Review E}\
  }\textbf {\bibinfo {volume} {87}},\ \bibinfo {pages} {065202} (\bibinfo
  {year} {2013})}\BibitemShut {NoStop}%
  \bibitem [{\citenamefont {Bardeen}(1991)}]{NLSsim1}%
  \BibitemOpen
  \bibfield  {author} {\bibinfo {author} {\bibfnamefont {S.}~\bibnamefont
  {Dyachenko}}, \bibfield  {author} {\bibinfo {author} {\bibfnamefont {A.C.}~\bibnamefont
  {Newell}}}, \bibfield  {author} {\bibinfo {author} {\bibfnamefont {A.}~\bibnamefont
  {Pushkarev}}},
  \bibinfo {author} {\bibfnamefont {V.E.}~\bibnamefont
  {Zakharov}}\ }\bibfield  {title} {\enquote {\bibinfo {title} {{Optical turbulence: weak turbulence, condensates and collapsing
filaments in the nonlinear Schrodinger equation}},}\ }\href@noop
  {} {\bibfield  {journal} {\bibinfo  {journal} {Physica D}\
  }\textbf {\bibinfo {volume} {57}},\ \bibinfo {pages} {96--160} (\bibinfo
  {year} {1992})}\BibitemShut {NoStop}%
  \bibitem [{\citenamefont {Bardeen}(1957)}]{NLSsim2}%
  \BibitemOpen
  \bibfield  {author} {\bibinfo {author} {\bibfnamefont {N.}~\bibnamefont
  {Vladimirova}}, \bibfield  {author} {\bibinfo {author} {\bibfnamefont {S.}~\bibnamefont
  {Derevyanko}}}, \bibfield  {author} {\bibinfo {author} {\bibfnamefont {F.}~\bibnamefont
  {Falkovich}}}\ }\bibfield  {title} {\enquote {\bibinfo {title} {{Phase transitions in wave turbulence}},}\ }\href@noop
  {} {\bibfield  {journal} {\bibinfo  {journal} {Physical Review E}\
  }\textbf {\bibinfo {volume} {85}},\ \bibinfo {pages} {010101(R)} (\bibinfo
  {year} {2012})}\BibitemShut {NoStop}%
\end{thebibliography}
\end{document}